\documentclass[conference]{IEEEtran}

\IEEEoverridecommandlockouts

\usepackage{cite}
\usepackage{amsmath,amssymb,amsfonts}
\usepackage{graphicx}
\usepackage{xcolor}
\usepackage{booktabs}
\usepackage[hidelinks]{hyperref}
\usepackage{textcase}

\title{TeRFS: Temporal-Evolving Radio Field Synthesis%
}

\author{
\IEEEauthorblockN{
Pengyang Zhang$^{1}$,
Wenlihan Lu$^{1}$,
Shijian Gao$^{1}$
}
\IEEEauthorblockA{$^{1}$
The Hong Kong University of Science and Technology (Guangzhou), Guangzhou, China}

\thanks{corresponding author: Shijian Gao. 
}
}

\begin{document}

\maketitle

\begin{abstract}
While radio-frequency (RF) field synthesis is fundamental to wireless networking, current approaches remain constrained by static assumptions, leaving them unable to track the rapid multipath reorganization of dynamic scenes. Modeling these transitions requires addressing two coupled challenges: explicit temporal representation and the capture of discrete path lifecycles. To bridge this gap, Temporal-Evolving Radio Field Synthesis (TeRFS) is introduced. TeRFS utilizes an anisotropic spherical Gaussian (ASG) directional basis to represent sparse, sharp angular structures, bound to analytical temporal envelopes that regulate path lifecycles. This formulation induces a mathematical birth-and-death mechanism, enabling individual multipath trajectories to emerge and vanish with temporal precision, a capability beyond the reach of standard smooth interpolation. Evaluations demonstrate that TeRFS outperforms state-of-the-art (SOTA) baselines, achieving an 11.5\% reduction in mean squared error (MSE) alongside a 6.9 times  training speedup. Even in environments characterized by extreme structural mutation, TeRFS maintains robust tracking of dynamic reorganizations, limiting median absolute error to 1.52 dB and establishing its utility for high-mobility wireless applications. The dataset and code is available at \href{https://github.com/zmydsg/TeRFS}{https://github.com/zmydsg/TeRFS}. 
\end{abstract}


\section{Introduction}
By providing site-specific characterizations of electromagnetic propagation, an accurate radio-frequency (RF) field serves as a physically consistent prior for 6G wireless~\cite{zeng2021ckm}, low-altitude networking~\cite{11417861}, and embodied navigation~\cite{liu2025embodiednavigation}. Traditional statistical models, though computationally efficient, lack 3D spatial awareness. On the other hand, differentiable ray tracing offers high precision but incurs prohibitive computational costs. Recent data-driven approaches using 2D image generation~\cite{wang2024radiodiff, gao2026farmfoundationalaerialradio} often fail to faithfully capture the underlying 3D radiated field due to this fundamental structural mismatch.

This limitation motivates representations that operate directly in 3D space. Building on neural radiance fields (NeRF), frameworks such as NeRF$^{2}$~\cite{zhao2023nerf2} utilize implicit volumetric rendering to parameterize scenes as continuous functions. Nevertheless, they often encounter high inference costs and dense querying requirements. To achieve greater efficiency, subsequent research transitions toward explicit representations, most notably 3D Gaussian Splatting (3DGS)~\cite{kerbl2023gaussiansplatting}. Representative works like WRF-GS~\cite{wen2025wrfgsplus} adapt the optical rasterization pipeline for antenna-side reception, while GSRF~\cite{yang2025gsrf} employs complex-valued Gaussians and Fourier-Legendre bases for phase-aware synthesis. Other specialized variants further extend these paradigms to handle path-level channel state information (CSI)~\cite{zhang2025rf3dgs}, outdoor radiomap extrapolation~\cite{wang2025radsplatter}, and real-time inference~\cite{zhang2025rfpgs}. Despite these architectural advancements, existing models share a fundamental limitation: they assume the scene remains static during acquisition. While some efforts address transceiver mobility~\cite{gao2025timevariant,quang2025dynamicrm,jia2025radiomapmotion,wen2025wrfgsplus, cao2025photonsplatting,chen2024rfcanvas,bian2025onewalk}, they typically rely on smooth variations in static environments and fail to model the continuous evolution of RF fields triggered by moving scatterers, which are, in essence, quasi-static assumptions.

This work addresses two primary challenges inherent in dynamic RF environments. The first is spatio-temporal parameterization, where moving beyond discrete frames requires extending 3D spatial representations into a continuous 4D manifold. To resolve this, the proposed  TeRFS establishes a continuous time axis by binding analytical 1D temporal envelopes to its spatial basis. The second, more fundamental challenge is structural multipath reorganization. Unlike optical scenes where motion causes localized deformation~\cite{wu2024_4dgs}, moving RF scatterers trigger global topological shifts by blocking existing trajectories and reflecting new ones. These abrupt birth-and-death events of propagation paths are poorly captured by globally supported bases like spherical harmonics (SH) or Fourier-Legendre Expansion (FLE)~\cite{yang2025gsrf}. Such bases act as low-pass filters and introduce ringing artifacts when attempting to fit sharp, sparse multipath peaks.

To capture these discontinuities, TeRFS utilizes an ASG directional basis~\cite{xu2013asg}. The choice of ASG is motivated by its compact local support, which aligns with the angular sparsity of RF multipaths and allows for high-precision modeling of narrow lobes. By integrating bounded temporal envelopes with individual ASG lobes, the framework formulates a birth-and-death mechanism. Unlike implicit time encoding which struggles with sudden changes, this analytical formulation allows specific multipath components to be mathematically activated or deactivated at precise timestamps. Consequently, the model tracks individual multipath trajectories as they emerge from reflections or vanish behind obstructions, maintaining a stable background while accurately rendering transient signals within a unified, continuous pipeline.

To evaluate TeRFS, a novel open-source dataset for temporal RF radiance field synthesis is constructed using a time-series ray-tracing engine within a 3D campus scene. Simulations involve multiple moving vehicles and a UAV to generate an evolving RF field characterized by severe multipath mutations. Results demonstrate that TeRFS reaches a mean peak signal-to-noise ratio (PSNR) of 19.45 dB while training 6.9 times faster than SOTA baselines. Crucially, for temporal reconstruction, TeRFS achieves a median absolute RSS error of 1.52 dB, which demonstrates a robust capability to track dynamic variations rather than passively averaging time-varying measurements.

\section[Architecture of TeRFS]{Architecture of T\textnormal{e}RFS}
\begin{figure*}[t]
    \centering
    \includegraphics[width=\textwidth]{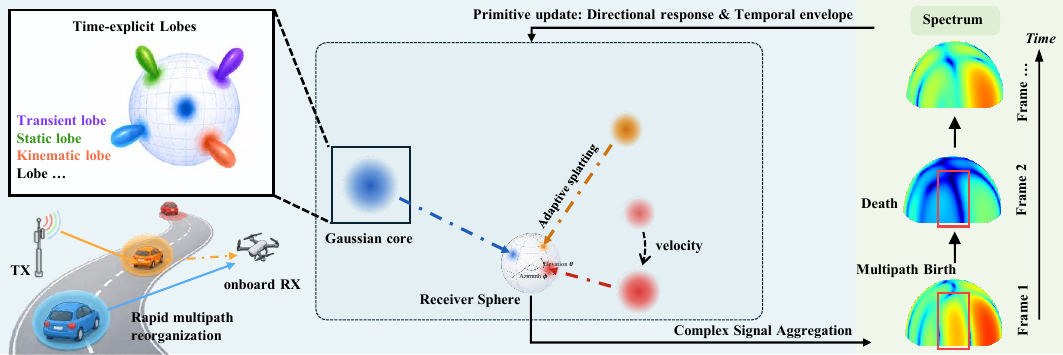}
    \caption{ System overview of TeRFS. In a dynamic urban scene where moving scatterers trigger rapid multipath reorganization, TeRFS parameterizes the propagation environment as a set of Gaussian primitives, each anchoring a localized propagation element at its center and carrying multiple ASG directional lobes. Every lobe is bound to a Gaussian temporal envelope with effective temporal support. Complementary static, kinematic, and transient behaviors represent persistence, motion, and short-lived events. Their temporal activation realizes the birth and death of individual multipaths observed in the frame-wise spectrum on the right. At each timestamp, active lobes are projected onto a receiver-centered sphere via adaptive splatting and aggregated in the complex domain into the predicted angular spectrum.}
    \label{fig:model_overview}
\end{figure*}

This section details the TeRFS architecture. We first formulate RF synthesis as a spatio-temporal reconstruction problem and then introduce the key system components. Finally, we describe the projection and aggregation of Gaussian components at the receiver. Fig. ~\ref{fig:model_overview} provides a system overview.
\subsection{Problem Statement}

Given an RF observation dataset $\mathcal{D}_{\mathrm{train}}
= \{(\mathbf{r}_i,t_i,\mathbf{S}_i)\}_{i=1}^{N}$
 where \(i\) indexes a spatio-temporal observation sample, \(\mathbf{r}_i\in\mathbb{R}^{3}\) is the receiver position, \(t_i\in\mathbb{R}\) is the timestamp, and \(\mathbf{S}_i\in\mathbb{R}^{H\times W}\) denotes the receiver-side angular power spectrogram with \(H\) elevation bins and \(W\) azimuth bins. The scalar received signal strength (RSS), denoted by \(s_i\), is obtained by converting $\mathbf{S}_i$ to linear scale, summing over all angular bins, and converting back to dB. The reported dB-domain RSS MSE and RSS absolute-error metrics are computed from this scalar RSS.

The objective is to learn a continuous TeRFS representation \(f_{\Theta}(\mathbf{x},t)\) that predicts the angular spectrogram
\(\hat{\mathbf{S}}_i=f_{\Theta}(\mathbf{r}_i,t_i)\) and further obtains the predicted RSS \(\hat{s}_i\). The learnable parameter set \(\Theta\) includes Gaussian geometry, directional lobe parameters, temporal envelope parameters, and rendering-related parameters. The reconstruction objective is written as
\begin{equation}
    \Theta^{*}
    =
    \arg\min_{\Theta}
    \sum_{i=1}^{N}
    \mathcal{L}
    \left(
    f_{\Theta}(\mathbf{r}_i,t_i),
    \mathbf{S}_i
    \right),
    \label{eq:objective}
\end{equation}
where \(\mathcal{L}\) denotes the training loss over the angular observation domain. To model time-evolving RF fields, \(f_{\Theta}\) should support both high-frequency spatial locality and time-dependent component activation. This allows the representation to capture the abrupt emergence and disappearance of multipath components while keeping the overall rendering process differentiable.

\subsection{Directional Basis Design}

To instantiate \(f_{\Theta}(\mathbf{x},t)\), TeRFS parameterizes the RF propagation scene as a set of 3D Gaussian primitives. Here, the scene denotes the physical propagation environment formed by background structures and scatterers, while each Gaussian primitive represents a learnable scattering or propagation element in this environment. The set of primitives is denoted as $\mathcal{G}
    =
    \{G_k\}_{k=1}^{K}$. 
Each primitive \(G_k\) is characterized by its center position \(\boldsymbol{\mu}_k\), covariance matrix \(\boldsymbol{\Sigma}_k\), base attenuation factor \(\rho_k\), and directional radiation response \(R_k\).

Standard neural radiance fields often use globally supported bases such as SH or FLE. In RF fields, however, multipath propagation usually forms sparse and sharp angular energy peaks. A global basis may spread local fitting errors to unrelated directions and cause ringing artifacts. To represent these localized angular components more compactly, we adopt the ASG basis, whose local support matches the angular sparsity of RF multipath propagation. Each primitive \(G_k\) is initialized with \(M\) = 32 ASG lobes to parameterize its directional response \(R_k\). For the \(m\)-th lobe of the \(k\)-th primitive, the learnable directional parameters are
\begin{equation}
    \{\mathbf{v}_{k,m},\boldsymbol{\lambda}_{k,m},
    \bar{a}_{k,m},\psi_{k,m}\},
\end{equation}
where \(\mathbf{v}_{k,m}\in\mathbb{S}^{2}\) is the lobe direction, \(\boldsymbol{\lambda}_{k,m}=(\lambda_x,\lambda_y)\) controls the anisotropic angular spread, \(\bar{a}_{k,m}\) is the base amplitude, and \(\psi_{k,m}\) is the carrier phase. For any query direction \(\mathbf{d}\), the directional response at time $t$ is
\begin{equation}
    R_k(\mathbf{d},t)
    =
    \sum_{m=1}^{M}
    a_{k,m}(t)
    e^{-j\psi_{k,m}}
    w_{k,m}(\mathbf{d}),
    \label{eq:directional_response}
\end{equation}
where \(a_{k,m}(t)\) is the time-dependent lobe amplitude defined in the next subsection. The ASG directional gating weight is
\begin{equation}
    w_{k,m}(\mathbf{d})
    =
    \exp
    \left(
    -
    \left(
    \lambda_x d_{\perp,x}^{2}
    +
    \lambda_y d_{\perp,y}^{2}
    \right)
    \right),
    \label{eq:asg_weight}
\end{equation}
where \(d_{\perp,x}\) and \(d_{\perp,y}\) are the components of \(\mathbf{d}\) on the local tangent plane of \(\mathbf{v}_{k,m}\). The lobe superposition forms the angular basis for temporal evolution.

\subsection{Time-Explicit Evolution}

After defining the directional lobes, we specify the time-dependent amplitude \(a_{k,m}(t)\). For each lobe, TeRFS uses a one-dimensional Gaussian temporal envelope parameterized by a temporal center \(t_c\) and an effective width \(t_w\):
\begin{equation}
    a_{k,m}(t)
    =
    \bar{a}_{k,m}
    \exp
    \left(
    -
    \frac{(t-t_c)^2}{2t_w^2}
    \right).
    \label{eq:temporal_envelope}
\end{equation}
We regard \([t_c-3t_w,t_c+3t_w]\) as the effective temporal support of the lobe.\footnote{This follows the common \(3\sigma\) criterion for Gaussian envelopes.} This envelope gives each directional lobe an explicit temporal activation range, so that different multipath components can evolve independently over time.

TeRFS represents persistent, motion-induced, and event-driven propagation through complementary static, kinematic, and transient lobe behaviors. Static behavior uses a large \(t_w\) for background paths that persist over the observed interval. Kinematic behavior follows \(\boldsymbol{\mu}_k(t)=\boldsymbol{\mu}_k(0)+\mathbf{u}_k t\), capturing gradual motion-induced path variation. Transient behavior uses a small \(t_w\), with \(t_c\) centered at the event time, to model short-lived components caused by occlusion, new reflections, or abrupt state changes. Together, these behaviors form a unified time-explicit representation of the evolving RF field.

\subsection{Adaptive Splatting}

Given the time-evolving Gaussian scattering components above, we next describe how their contributions are projected to the receiver-side angular domain and aggregated into the predicted angular spectrogram. Since the RF domain does not have a physical 2D imaging plane, TeRFS constructs a virtual observation sphere with radius \(r_{\mathrm{rx}}\) centered at the receiver position \(\mathbf{r}\). For a point \((x,y,z)\) on this sphere, the azimuth \(\phi\) and elevation \(\theta\) are defined as
\begin{equation}
    \phi
    =
    \operatorname{atan2}(y,x),
    \qquad
    \theta
    =
    \frac{\pi}{2}
    -
    \arccos
    \left(
    \frac{z}{r_{\mathrm{rx}}}
    \right).
    \label{eq:spherical_coordinates}
\end{equation}

Let \(\mathbf{J}_k\) be the Jacobian of the spherical projection from 3D position to angular coordinates \((\phi,\theta)\), evaluated at \(\boldsymbol{\mu}_k\). The 3D covariance is projected onto the 2D angular domain as $    \boldsymbol{\Sigma}^{2D}_k
    =
    \mathbf{J}_k
    \boldsymbol{\Sigma}_k
    \mathbf{J}_k^{\top}$. 
A purely geometric projection may underestimate the angular coverage of nearby scatterers. Following the near-field intuition behind the Fraunhofer criterion, TeRFS applies a distance-aware covariance compensation:
\begin{equation}
    \boldsymbol{\Sigma}^{\prime 2D}_k
    =
    \boldsymbol{\Sigma}^{2D}_k
    \left(
    1+\frac{\eta^2}{D_k^2}
    \right),
    \label{eq:adaptive_covariance}
\end{equation}
where \(D_k=\|\boldsymbol{\mu}_k-\mathbf{r}\|\) is the distance from the primitive center to the receiver. Here, \(\eta\) is a jointly optimized global scale parameter. We use a single global value to keep the compensation tied to scene-level near-field spreading and avoid direction-specific overfitting. When \(D_k\) is large, the compensation term becomes small and the projection approaches the original geometric projection. After projection, the effective coverage radius of each primitive on the angular grid is determined by $    r_k
    =
    3
    \sqrt{
    \lambda_{\max}
    \left(
    \boldsymbol{\Sigma}^{\prime 2D}_k
    \right)
    }
    \label{eq:coverage_radius}$. For an angular grid point \(\mathbf{q}_j\), the opacity weight of the \(k\)-th primitive is
\begin{equation}
    \alpha_k(\mathbf{q}_j)
    =
    \rho_k
    \exp
    \left(
    -
    \frac{1}{2}
    \Delta\mathbf{q}_k^{\top}
    \left(
    \boldsymbol{\Sigma}^{\prime 2D}_k
    \right)^{-1}
    \Delta\mathbf{q}_k
    \right),
    \label{eq:opacity_weight}
\end{equation}
where \(\Delta\mathbf{q}_k=\mathbf{q}_j-\boldsymbol{\mu}^{2D}_k\) is the angular deviation from the projected center \(\boldsymbol{\mu}^{2D}_k\). All primitives covering \(\mathbf{q}_j\) are depth-sorted according to \(D_k\) and accumulated in the complex domain. The accumulated transmittance is
\begin{equation}
    T_k
    =
    \prod_{\ell<k}
    \left(
    1-\alpha_{\ell}(\mathbf{q}_j)
    \right),
    \label{eq:transmittance}
\end{equation}
where \(\ell<k\) follows the depth-sorted order. The complex received response at angular bin \(\mathbf{q}_j\) and timestamp \(t\) is
\begin{equation}
    z_{t,j}
    =
    \sum_{k}
    T_k
    \alpha_k(\mathbf{q}_j)
    R_k(\mathbf{d}_{k,j},t),
    \label{eq:complex_aggregation}
\end{equation}
where \(\mathbf{d}_{k,j}\) is the observation direction associated with the \(k\)-th primitive and angular bin \(\mathbf{q}_j\). The predicted RSS \(\hat{s}_t\) is obtained by converting the bin-wise complex responses to angular power values and then integrating them over the angular bins.

\section[Implementation of TeRFS]{Implementation of T{\textnormal{e}}RFS}

This section presents the TeRFS implementation in two parts: we define the spatio-temporal training objectives used to optimize the representation and describe the birth-and-death mechanism that updates path events during optimization.

\subsection{Spatio-Temporal Training Objectives}

In each training iteration, we sample a batch of \(B\) timestamps \(\{t_b\}_{b=1}^{B}\) from the temporal domain. The total training loss is
\begin{equation}
    \mathcal{L}
    =
    \mathcal{L}_{\mathrm{recon}}
    +
    \lambda_{\mathrm{td}}
    \mathcal{L}_{\mathrm{TD}}
    +
    \lambda_{\mathrm{gate}}
    \mathcal{H}_{\mathrm{gate}},
    \label{eq:total_loss}
\end{equation}
where \(\mathcal{L}_{\mathrm{recon}}\) supervises angular reconstruction, \(\mathcal{L}_{\mathrm{TD}}\) constrains temporal variation, and \(\mathcal{H}_{\mathrm{gate}}\) encourages sparse temporal lobe activation.

\textbf{Reconstruction loss:} 
At each receiver location and timestamp, TeRFS predicts an angular power spectrogram \(\hat{\mathbf{S}}_{t_b}\). The reconstruction loss is applied on this angular representation so that the model retains directional multipath structure:
\begin{equation}
    \mathcal{L}_{\mathrm{recon}}
    =
    (1-\lambda_1-\lambda_2)\mathcal{L}_{1}
    +
    \lambda_1\mathcal{L}_{\mathrm{SSIM}}
    +
    \lambda_2\mathcal{L}_{\mathrm{Fourier}}
    \label{eq:recon_loss}
\end{equation}
Here, \(\mathcal{L}_{1}\) measures point-wise amplitude error, \(\mathcal{L}_{\mathrm{SSIM}}\) encourages structural consistency on the angular grid, and \(\mathcal{L}_{\mathrm{Fourier}}\) applies frequency-domain weighting to fit sharp angular energy peaks induced by sparse multipath components.

\textbf{Temporal difference loss:} 
Adjacent-frame smoothing may suppress valid abrupt changes caused by moving scatterers. Instead, TeRFS matches relative changes between sampled timestamps:
\begin{equation}
    \mathcal{L}_{\mathrm{TD}}
    =
    \frac{1}{B(B-1)}
    \sum_{b=1}^{B}\sum_{\substack{b'=1 \\ b'\neq b}}^{B}
    \left\|
    \left(
    \hat{\mathbf{S}}_{t_b}
    -
    \hat{\mathbf{S}}_{t_{b'}}
    \right)
    -
    \left(
    \mathbf{S}_{t_b}
    -
    \mathbf{S}_{t_{b'}}
    \right)
    \right\|_{1}
    \label{eq:td_loss}
\end{equation}
This loss constrains temporal variation in the angular spectrogram while still allowing event-like changes in the multipath structure.

\textbf{Lobe-gate entropy regularization:} 
To avoid keeping all lobes weakly active over the entire sequence, we encourage sparse temporal activation among the lobes. Let
\(\xi_{k,m}(t_b)= (t_b-t_{c,k,m})^2/(2t_{w,k,m}^2)\)
denote the temporal activation score of the \(m\)-th lobe of primitive \(G_k\). The normalized temporal activation~is
\begin{equation}
    \pi_{k,m}(t_b)
    =
    \frac{
    \exp\!\left(-\xi_{k,m}(t_b)\right)
    }{
    \sum_{m'=1}^{M}
    \exp\!\left(-\xi_{k,m'}(t_b)\right)
    }
    \label{eq:lobe_activation}
\end{equation}
The lobe-gate entropy is
\begin{equation}
    \mathcal{H}_{\mathrm{gate}}
    =
    -
    \frac{1}{BK}
    \sum_{b=1}^{B}
    \sum_{k=1}^{K}
    \sum_{m=1}^{M}
    \pi_{k,m}(t_b)
    \log
    \pi_{k,m}(t_b).
    \label{eq:gate_entropy}
\end{equation}
Minimizing this term encourages more decisive temporal activation among the lobes.

\subsection{Birth-and-Death Mechanism}

The Gaussian temporal envelope is differentiable with respect to \(t_c\) and \(t_w\). Applying the chain rule to \(a_{k,m}(t)\) gives
\begin{subequations}
\label{eq:temporal_gradients}
\begin{align}
    \frac{\partial \mathcal{L}}{\partial t_c}
    &=
    \frac{\partial \mathcal{L}}{\partial a_{k,m}(t)}
    a_{k,m}(t)
    \frac{t-t_c}{t_w^2},
    \label{eq:grad_tc}
    \\
    \frac{\partial \mathcal{L}}{\partial t_w}
    &=
    \frac{\partial \mathcal{L}}{\partial a_{k,m}(t)}
    a_{k,m}(t)
    \frac{(t-t_c)^2}{t_w^3}.
    \label{eq:grad_tw}
\end{align}
\begin{figure}[!t]
    \centering
    \includegraphics[width=0.76\linewidth]{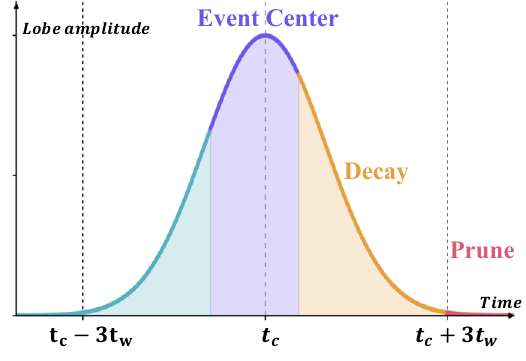}
    \caption{Lobe-wise temporal gate. Each lobe is active within $[t_c-3t_w,t_c+3t_w]$: it is born when a primitive's gradient signals an event, peaks at $t_c$, and becomes inactive outside its temporal support. Inactive transient lobes are pruned during optimization.}
    \label{fig:envolope}
\end{figure}
\end{subequations}
These gradients adapt \(t_c\) to the event time and \(t_w\) to its duration. The mechanism further incorporates discrete lobe birth and pruning.

\textbf{Gradient-driven transient birth:} 
Unlike visual 3DGS, where positional gradients mainly indicate geometric misalignment, abrupt RF changes often appear as local angular energy changes. TeRFS tracks the amplitude-gradient score for each existing lobe:
\begin{equation}
    \Gamma_{k,m}
    =
    \sum_{b=1}^{B}
    \left|
    \frac{\partial \mathcal{L}}
    {\partial a_{k,m}(t_b)}
    \right|.
\end{equation}

When \(\Gamma_{k,m}\) exceeds a preset threshold \(\tau\), a transient lobe is created, as illustrated in Fig. \ref{fig:envolope}. Its temporal center is initialized as \(t_c=t_{b^\star}\), where \(b^\star=\arg\max_b |\partial\mathcal L/\partial a_{k,m}(t_b)|\). Its width is initialized as \(t_w=\Delta t\), the minimum temporal width set by the sampling interval. This update introduces a short-lived lobe when the current representation cannot explain a local temporal change.

\textbf{Attenuation-driven transient pruning:} At a later queried timestamp \(t'\), if
\begin{equation}
    t'
    \notin
    [t_c-3t_w,\; t_c+3t_w],
\end{equation}
the corresponding transient event is treated as inactive and can be pruned. Physically, this corresponds to a propagation path disappearing due to occlusion, the scatterer leaving the relevant propagation geometry, or loss of angular support. This pruning limits memory growth during long-sequence optimization while preserving short-lived multipath~events.

\section{Experiments}

\subsection{Dataset and Experimental Setup}

\begin{figure}[t]
    \centering
    \includegraphics[width=0.8\linewidth]{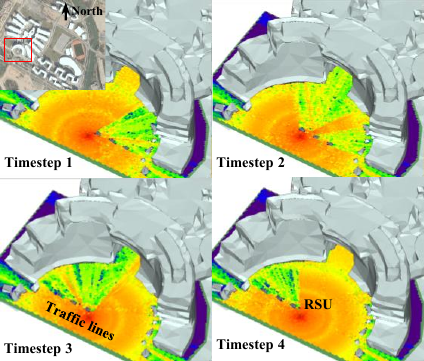}
    \caption{Four partial snapshots of the dataset, where vehicles passing by the RSU induce drastic reorganization of the RF field across timesteps.}
    \label{fig:dataset}
\end{figure}

To evaluate spatial RF field synthesis and temporal multipath evolution, we construct the \textit{TeRFS-HKUST (GZ)} dataset. As illustrated in Fig. ~\ref{fig:dataset}, the dataset is generated in a campus building-and-road scene at HKUST-GZ using the Wireless InSite 4.0 ray tracer, with up to 6 reflections, no diffractions, and 1 transmission. A single Roadside Unit (RSU) transmitter operates at 5.9 GHz with 20 dBm transmit power. The temporal sequence spans 10.0 seconds at 10 Hz, resulting in 100 frames. Seven moving scatterers, including six vehicles and one UAV, pass close to the RSU and create strong time-varying perturbations in the RF field. The receiver grid contains 2,008 effective receivers placed on a 3 m horizontal grid at 1 m height. Each receiver is equipped with a \(4\times4\) planar array, and the resulting angular observations are normalized to the \([-100,-40]\) dBm range.

TeRFS was evaluated from two perspectives. For single-frame spatial synthesis, we use the initial frame and split the 2,008 receivers into 80\% training and 20\% testing receivers. We compare TeRFS with three representative baselines: NeRF\(^2\)~\cite{zhao2023nerf2}, an implicit RF radiance-field method; WRF-GS+~\cite{wen2025wrfgsplus}, an explicit Gaussian-splatting RF baseline; and GSRF~\cite{yang2025gsrf}, a fully explicit complex-valued Gaussian baseline. For temporal interpolation, we use the 100-frame sequence with a 70/30 random train/test split over time. Since the baselines do not include an explicit temporal parameterization, temporal interpolation is evaluated for TeRFS only. All models are trained using Adam on a single RTX 4090 GPU. TeRFS uses a three-stage schedule: static warm-up on the initial frame, lobe-level temporal gate optimization, and final joint fine-tuning.

\subsection{Spatial Generalization in Static Scenes}

\begin{figure}[t]
    \centering
    \includegraphics[width=\columnwidth]{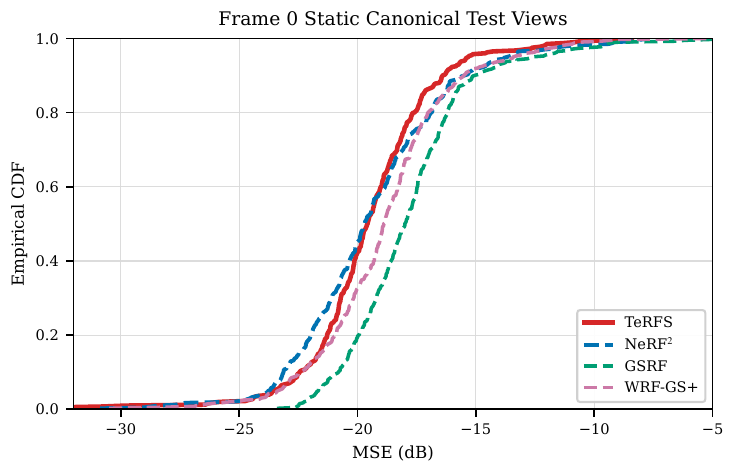}
    \caption{Empirical cumulative distribution function (ECDF) of the MSE for single-frame spatial synthesis over 402 unseen test receivers.}
    \label{fig:mse_cdf}
\end{figure}

In the single-frame setting, TeRFS achieves spatial synthesis quality comparable with the strongest baselines while using a more efficient explicit representation. As shown in Fig. ~\ref{fig:mse_cdf}, TeRFS obtains the lowest mean MSE among all methods. It reduces the mean MSE by 11.5\% over NeRF\(^2\), and by 27.8\% over the two explicit baselines on average. Its median MSE is also close to NeRF\(^2\) and lower than both explicit baselines. Despite a minor performance drop compared to NeRF² in smooth, low-MSE regions, TeRFS excels in complex, high-error cases (the top 10\% error bracket), delivering an average MSE reduction exceeding 25\% at challenging receiver locations. Its mean PSNR of 19.45 dB remains competitive with existing baselines on static reconstruction quality. This accuracy is achieved without dense volumetric querying, as the ASG directional lobes provide a compact local basis for sparse angular multipath components. As a sparse-data check, TeRFS still reaches 17.53 dB PSNR when trained with only 25\% of the training receivers.

\subsection{Temporal Reconstruction in Dynamic Scenes}

\begin{figure}[t]
    \centering
    \includegraphics[width=\columnwidth]{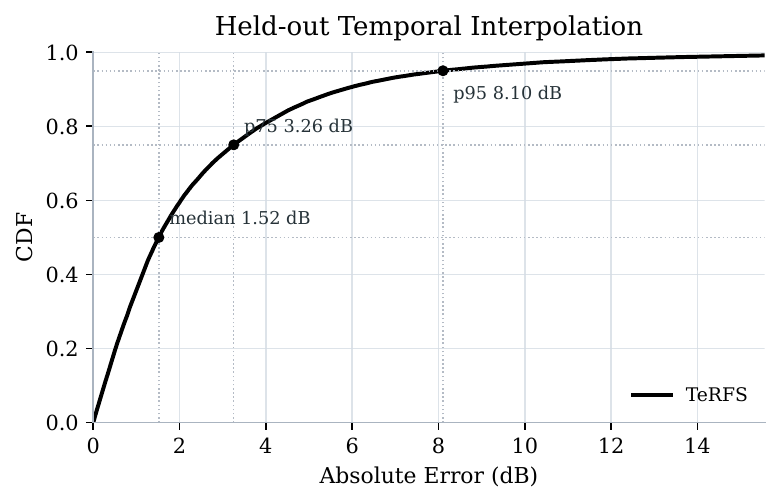}
    \caption{Empirical CDF of absolute RSS error across 60,240 temporal test samples.}
    \label{fig:temporal_cdf}
\end{figure}

For temporal interpolation, TeRFS is evaluated on 60,240 unseen spatio-temporal samples. It achieves a mean RSS absolute error of 2.53 dB and a median error of 1.52 dB, with 75\% of the samples below 3.3 dB. As shown in Fig. ~\ref{fig:temporal_cdf}, the time-explicit lobe representation interpolates unseen frames with low RSS error.

\begin{figure}[t]
    \centering
    \includegraphics[width=1.0\columnwidth]{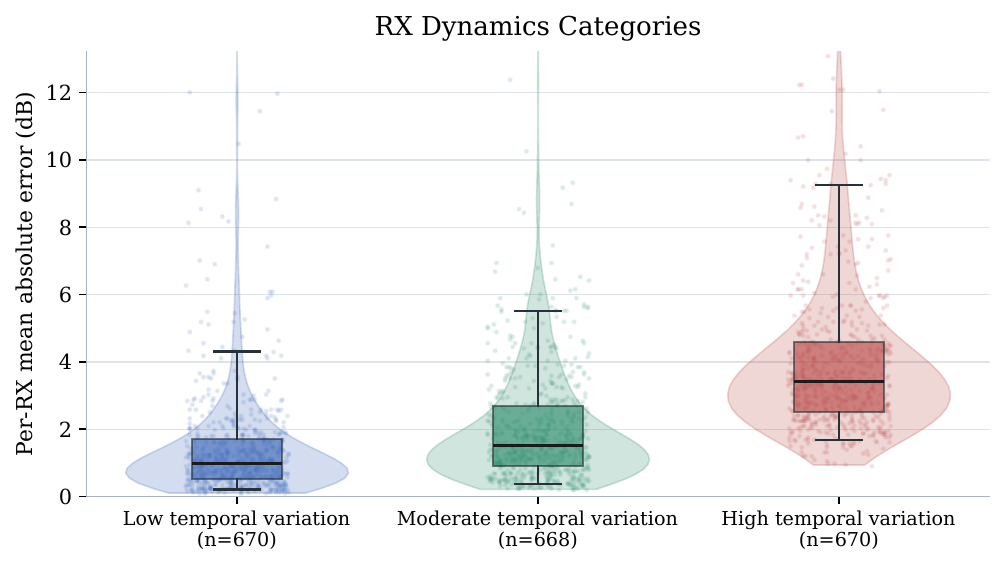}
    \caption{Error under increasing receiver dynamics. Test receivers are grouped by the temporal standard deviation of their ground-truth RSS.}
    \label{fig:rx_dynamics_stratified}
\end{figure}

We further group test receivers into low, moderate, and high dynamic tiers according to the temporal standard deviation of their ground-truth RSS. As shown in Fig. ~\ref{fig:rx_dynamics_stratified}, the mean / median errors are 1.44 / 0.99 dB for the low-dynamic tier, 2.04 / 1.39 dB for the moderate-dynamic tier, and 4.11 / 2.81 dB for the high-dynamic tier. In the most dynamic tier, the ground-truth RSS varies with a 5.29 dB temporal standard deviation, while the median error remains 2.81 dB. This shows that the prediction error grows much more slowly than the underlying RSS variation, rather than collapsing to a temporal average. Rare large errors still appear under near-instantaneous occlusion events, where the current continuous temporal envelopes may smooth abrupt transitions. 

\subsection{Computational Complexity and Representation Efficiency}

\begin{figure}[!htbp]
    \centering
    \includegraphics[width=\columnwidth]{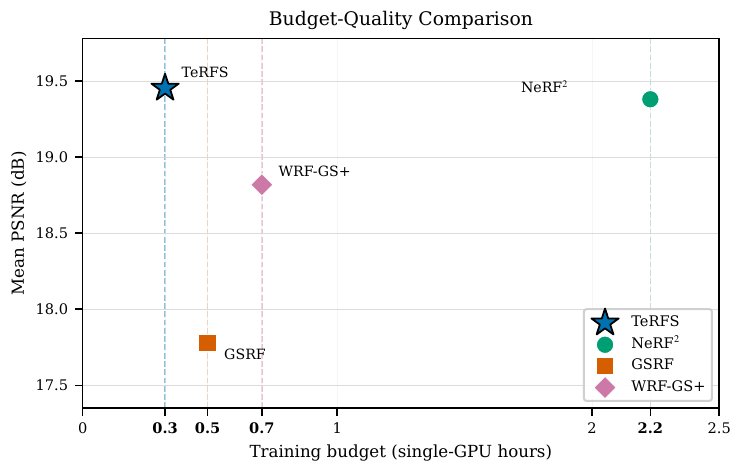}
    \caption{Training time versus mean PSNR under default training schedules. TeRFS achieves a favorable quality-efficiency trade-off.}
    \label{fig:pareto}
\end{figure}

Finally, we evaluate the quality-efficiency trade-off of TeRFS. As shown in Fig. ~\ref{fig:pareto}, TeRFS reaches competitive mean PSNR with a much lower training budget. It requires approximately 0.32 hours of training, running 6.9 times faster than NeRF\(^2\), which takes 2.2 hours.

For a single \(90\times360\) view with batch size 1, TeRFS runs in 3.11 ms with 273 MB peak VRAM in the single-frame setting. Temporal inference also remains fast at 4.16 ms, with 109 MB peak VRAM. Compared with explicit baselines, TeRFS uses only 36.9\% of the primitives, showing that ASG directional lobes provide a compact representation for sparse RF multipath~structure.

\section{Conclusions}
This work introduced TeRFS, a novel spatio-temporal representation that elevated radio field synthesis from static spatial reconstructions to dynamic spatio-temporal domains. TeRFS couples analytical temporal envelopes with an ASG directional basis to implement a birth-and-death mechanism. In contrast to existing smooth-interpolation models, this formulation effectively captured discontinuous, global topological reorganizations of RF events. Empirical evaluations confirmed that TeRFS maintained superior reconstruction fidelity with significant acceleration over existing baselines. Furthermore, the framework demonstrated robust temporal interpolation, tracking structural mutations that typically cause implicit models to fail. TeRFS establishes a new foundation for high-fidelity sensing in non-stationary wireless environments. Future research will focus on extending this framework toward temporal extrapolation and long-term predictive modeling for ultra-mobile communication~systems.

\bibliographystyle{IEEEtran}
\bibliography{references}

@ARTICLE{11417861,
  author={Gao, Shijian and Yan, Jia and Huang, Peiyuan and Lu, Ziyi and Gong, Mingze and Miao, Linghui and Zhu, Gangyong and Liang, Jiahui and Yang, Liuqing},
  journal={IEEE Internet of Things Magazine}, 
  title={Integrated Sensing, Communication, and Computation for Low-Altitude Networks Towards Seamless Connectivity and Connected Intelligence}, 
  year={2026},
  volume={9},
  number={3},
  pages={63-71},
  doi={10.1109/MIOT.2026.3660848}}

@inproceedings{zhao2023nerf2,
  author    = {Zhao, Xiaopeng and An, Zhenlin and Pan, Qingrui and Yang, Lei},
  title     = {{{NeRF$^2$}: Neural Radio-Frequency Radiance Fields}},
  booktitle = {Proceedings of the 29th Annual International Conference on Mobile Computing and Networking (MobiCom)},
  pages     = {1--15},
  year      = {2023},
  doi       = {10.1145/3570361.3592527}
}

@article{wen2025wrfgsplus,
  author  = {Wen, Chaozheng and Tong, Jingwen and Hu, Yingdong and Lin, Zehong and Zhang, Jun},
  title   = {{Neural Representation for Wireless Radiation Field Reconstruction: A {3D} Gaussian Splatting Approach}},
  journal = {IEEE Transactions on Wireless Communications},
  year    = {2025},
  doi     = {10.1109/TWC.2025.3631663}
}

@article{zhang2025rf3dgs,
  author  = {Zhang, Lihao and Sun, Haijian and Berweger, Samuel and Gentile, Camillo and Hu, Rose Qingyang},
  title   = {{{RF-3DGS}: Wireless Channel Modeling with Radio Radiance Field and {3D} Gaussian Splatting}},
  journal = {IEEE Transactions on Wireless Communications},
  year    = {2025},
  doi     = {10.1109/TWC.2024.3514919}
}

@inproceedings{yang2025gsrf,
  author    = {Yang, Kang and Dong, Gaofeng and Ji, Sijie and Du, Wan and Srivastava, Mani},
  title     = {{{GSRF}: Complex-Valued {3D} Gaussian Splatting for Efficient Radio-Frequency Data Synthesis}},
  booktitle = {Proceedings of the 39th Conference on Neural Information Processing Systems (NeurIPS)},
  year      = {2025}
}

@article{cao2025photonsplatting,
  author  = {Cao, Ge and Gradoni, Gabriele and Peng, Zhen},
  title   = {{Photon Splatting: A Physics-Guided Neural Surrogate for Real-Time Wireless Channel Prediction}},
  journal = {arXiv preprint arXiv:2507.04595},
  year    = {2025}
}

@article{wang2025radsplatter,
  author  = {Wang, Yiheng and Xue, Ye and Zhang, Shutao and Chang, Tsung-Hui},
  title   = {{{RadSplatter}: Extending {3D} Gaussian Splatting to Radio Frequencies for Wireless Radiomap Extrapolation}},
  journal = {arXiv preprint arXiv:2502.12686},
  year    = {2025}
}

@article{zhang2025rfpgs,
  author  = {Zhang, Lihao and Li, Zongtan and Sun, Haijian},
  title   = {{{RF-PGS}: Fully-Structured Spatial Wireless Channel Representation with Planar Gaussian Splatting}},
  journal = {arXiv preprint arXiv:2508.16849},
  year    = {2025}
}

@article{bian2025onewalk,
  author  = {Bian, Yiheng and others},
  title   = {{One Walk is All You Need: Data-Efficient {3D} {RF} Scene Reconstruction with Human Movements}},
  journal = {arXiv preprint arXiv:2511.16966},
  year    = {2025}
}

@article{wang2024radiodiff,
  author  = {Wang, Xiucheng and others},
  title   = {{{RadioDiff}: An Effective Generative Diffusion Model for Sampling-Free Dynamic Radio Map Construction}},
  journal = {IEEE Transactions on Cognitive Communications and Networking},
  volume  = {11},
  number  = {2},
  pages   = {738--750},
  year    = {2025},
  doi     = {10.1109/TCCN.2024.3504489}
}

@inproceedings{chen2024rfcanvas,
  author    = {Chen, Xingyu and Feng, Zihao and Sun, Ke and Qian, Kun and Zhang, Xinyu},
  title     = {{{RFCanvas}: Modeling {RF} Channel by Fusing Visual Priors and Few-shot {RF} Measurements}},
  booktitle = {Proceedings of the 22nd ACM Conference on Embedded Networked Sensor Systems (SenSys)},
  pages     = {464--477},
  year      = {2024},
  doi       = {10.1145/3666025.3699351}
}

@article{gao2025timevariant,
  author  = {Gao, Qianhao and others},
  title   = {{Time-Variant Radio Map Reconstruction with Optimized Distributed Sensors in Dynamic Spectrum Environments}},
  journal = {IEEE Internet of Things Journal},
  volume  = {12},
  number  = {12},
  pages   = {20927--20941},
  year    = {2025},
  doi     = {10.1109/JIOT.2025.3545542}
}

@article{jia2025radiomapmotion,
  author  = {Jia, Honggang and Cheng, Nan and Wang, Xiucheng},
  title   = {{{RadioMapMotion}: A Dataset and Baseline for Proactive Spatio-Temporal Radio Environment Prediction}},
  journal = {arXiv preprint arXiv:2511.17526},
  year    = {2025}
}

@article{quang2025dynamicrm,
  author  = {Quang, Nguyen Duc Minh and Liu, Chang and Nguyen, Huy-Trung and Li, Shuangyang and Ng, Derrick Wing Kwan and Xiang, Wei},
  title   = {{{3D} Dynamic Radio Map Prediction Using Vision Transformers for Low-Altitude Wireless Networks}},
  journal = {arXiv preprint arXiv:2511.19019},
  year    = {2025}
}

@article{kerbl2023gaussiansplatting,
  author  = {Kerbl, Bernhard and Kopanas, Georgios and Leimk{\"u}hler, Thomas and Drettakis, George},
  title   = {{{3D} Gaussian Splatting for Real-Time Radiance Field Rendering}},
  journal = {ACM Transactions on Graphics},
  volume  = {42},
  number  = {4},
  pages   = {139:1--139:14},
  year    = {2023}
}

@inproceedings{wu2024_4dgs,
  author    = {Wu, Guanjun and others},
  title     = {{{4D} Gaussian Splatting for Real-Time Dynamic Scene Rendering}},
  booktitle = {Proceedings of the IEEE/CVF Conference on Computer Vision and Pattern Recognition (CVPR)},
  pages     = {20310--20320},
  year      = {2024}
}

@article{xu2013asg,
  author  = {Xu, Kun and Sun, Wei-Lun and Dong, Zhao and Zhao, Dan-Yong and Wu, Run-Dong and Hu, Shi-Min},
  title   = {{Anisotropic Spherical Gaussians}},
  journal = {ACM Transactions on Graphics},
  volume  = {32},
  number  = {6},
  pages   = {209:1--209:11},
  year    = {2013},
  doi     = {10.1145/2508363.2508386}
}

@article{gao2026farmfoundationalaerialradio,
  author  = {Gao, Shijian and Liang, Jiahui and Yuan, Yifeng and Lu, Wenlihan and Shen, Guobin and Yang, Liuqing},
  title   = {{{FARM}: Foundational Aerial Radio Map for Intelligent Low-Altitude Networking}},
  journal = {arXiv preprint arXiv:2604.17362},
  year    = {2026}
}

@article{zeng2021ckm,
  author  = {Zeng, Yong and Xu, Xiaoli},
  title   = {{Toward Environment-Aware 6G Communications via Channel Knowledge Map}},
  journal = {IEEE Wireless Communications},
  volume  = {28},
  number  = {3},
  pages   = {84--91},
  year    = {2021},
  doi     = {10.1109/MWC.001.2000327}
}

@article{liu2025embodiednavigation,
  author  = {Liu, Yunhao and Liu, Li and Zheng, Yawen and Liu, Yunhuai and Dang, Fan and Li, Ningbo and Ma, Ke},
  title   = {{Embodied Navigation}},
  journal = {Science China Information Sciences},
  volume  = {68},
  pages   = {141101},
  year    = {2025},
  doi     = {10.1007/s11432-024-4303-8}
}

\end{document}